%% file: JointSpk.tex
\documentclass[a4paper]{article}

\usepackage{INTERSPEECH2021}
\usepackage[percent]{overpic}
\usepackage{multirow}
\usepackage{hyperref}
\usepackage{xcolor}

\include{Notations}

\newcommand{\comment}[1]{}

\title{Cross-Lingual Text-to-Speech Using Multi-Task Learning and Speaker Classifier Joint Training}
\name{Jingzhou Yang and Lei He}
\address{Microsoft Azure Speech}
\email{\{jingy,helei\}@microsoft.com}

\begin{document}

\maketitle
\begin{abstract}
In cross-lingual speech synthesis, the speech in various languages can be synthesized for a monoglot speaker. Normally, only the data of monoglot speakers are available for model training, thus the speaker similarity is relatively low between the synthesized cross-lingual speech and the native language recordings. Based on the multilingual transformer text-to-speech model, this paper studies a multi-task learning framework to improve the cross-lingual speaker similarity. To further improve the speaker similarity, joint training with a speaker classifier is proposed. Here, a scheme similar to parallel scheduled sampling is proposed to train the transformer model efficiently to avoid breaking the parallel training mechanism when introducing joint training. By using multi-task learning and speaker classifier joint training, in subjective and objective evaluations, the cross-lingual speaker similarity can be consistently improved for both the seen and unseen speakers in the training set.
%
%
\end{abstract}

\section{Introduction}
\label{sec:intro}
In recent years, end-to-end (E2E) models have been widely used in 
speech synthesis~\cite{Sotelo:char2wav:2017,Arik:DeepVoice:2017,Taigman:VoiceLoop:2018,Wang:Tacotron:2018,Shen:Tacotron2:2018,Li:TranTTS:2019}, 
where the models can be directly trained on text-speech pairs with minimal engineering efforts.
Based on the encoder-decoder E2E framework, various multilingual text-to-speech (TTS) approaches have been 
proposed~\cite{Baljekar:MultiIndia:2018,Li:UTF8:2019,Zhang:Multi:2019,Nachmani:Polyglot:2019,Liu:crosslingual:2019,Yang:UniTTS:2020,Chen:Cross:2019,Cai:Cross:2020}. In these approaches, the speaker and language\footnote{The language embedding is optional for some systems.} embeddings are introduced to characterize the voice of the speaker, and the global prosody of the language. By using different combinations of the input sequence, speaker and language embeddings, cross-lingual synthesis can be achieved. 

Normally, only the data from monoglot speakers in different languages are available in multilingual model training. Without using polyglot data or any constrain in training, the generated cross-lingual speech can have low speaker similarity. This phenomenon has been exhibited in the previous work~\cite{Zhang:Multi:2019,Nachmani:Polyglot:2019,Yang:UniTTS:2020}. However, the approaches to improve cross-lingual speaker similarity are scarcely discussed. In~\cite{Zhang:Multi:2019}, a speaker classifier is introduced under the domain-adversarial training framework to encourage the encoded phone sequence to be speaker-independent. This approach, to some extent, improves the cross-lingual speaker similarity when there is only one or very few speakers per language in the training data. 
In~\cite{Nachmani:Polyglot:2019}, a polyglot loss is introduced to minimize the L1 distance between the speaker embeddings of the original recording and cross-lingual speech for the same speaker. This approach can only be applied to the systems using speaker encoders, which map the input speech sequence to a fixed-dimensional vector. For the systems using lookup tables, only a fixed or trainable embedding vector is associated with one speaker, then the polyglot loss becomes inapplicable. 

Our previous work on multilingual TTS~\cite{Yang:UniTTS:2020} shows that the multilingual model can achieve very good speaker similarity in intra-lingual synthesis, which is very close to the original recording. Thus, this paper focuses on improving the speaker similarity in cross-lingual scenarios. 

In this work, the multilingual model is based on the transformer TTS~\cite{Li:TranTTS:2019} by introducing additional speaker and language networks.
Based on the multilingual transformer TTS, a multi-task learning (MTL) framework is introduced to encourage the speaker and language embeddings to capture the speaker and language characteristics. Experiments show that MTL enhances the speaker representation and improves speaker similarity in cross-lingual synthesis. To further improve the cross-lingual speaker similarity, an x-vector speaker classifier~\cite{Snyder:Xvec:2018} is introduced to be jointly trained with the multilingual transformer\footnote{Although the transformer-based model is discussed, the general MTL and joint training frameworks can also be applied to other E2E multilingual models.}. In joint training, the distance between the x-vectors of the recording and the cross-lingual speech is minimized. As cross-lingual speech needs to be generated in jointing training, the transformer needs to be operated in inference mode, thus the model cannot be trained in parallel. To alleviate this problem, an approximate approach similar to parallel scheduled sampling~\cite{Duckworth:PSS:2019,Zhou:PSS:2019} is introduced to train the transformer in parallel.

The paper is organized as follows. In section~\ref{sec:framework} the general
framework of MTL and speaker classifier joint training is introduced. Experiments and corresponding evaluation results
are presented in section~\ref{sec:exp}. Finally, conclusions and the future
work are discussed in section~\ref{sec:conc}.

\comment{
However, the cross-lingual speaker similarity reported in these work are relatively low, and the approaches to improve similarity are scarcely discussed. In [], a cycle loss is introduced to improve the speaker similarity of the system using speaker encoder, which maps the input speech sequence to a fixed-dimensional vector. However, for the systems without using speaker encoder [], this approach is not applicable. In our previous multilingual work, the multilingual model can achieve very good speaker similarity which is very close to the recordings. However, the cross-lingual speaker similarity is relatively low, thus this work will focus on improving the cross-lingual speaker similarity.
}

\section{The general framework}
\label{sec:framework}
The multilingual transformer TTS~\cite{Yang:UniTTS:2020} is an extension 
of the transformer TTS~\cite{Li:TranTTS:2019} by introducing the speaker and
language conditions to the model. These conditions represent the speaker and
language global characteristics, and play key roles in multilingual modeling.
However, without using polyglot data or any constrain in training, such model
cannot guarantee a good speaker similarity in cross-lingual synthesis. 
To improve the cross-lingual speaker similarity, MTL and speaker classifier joint training are discussed in this section. 

\begin{figure}[t!]
	\begin{center}
		\begin{overpic}[grid=false,width=8.2cm]{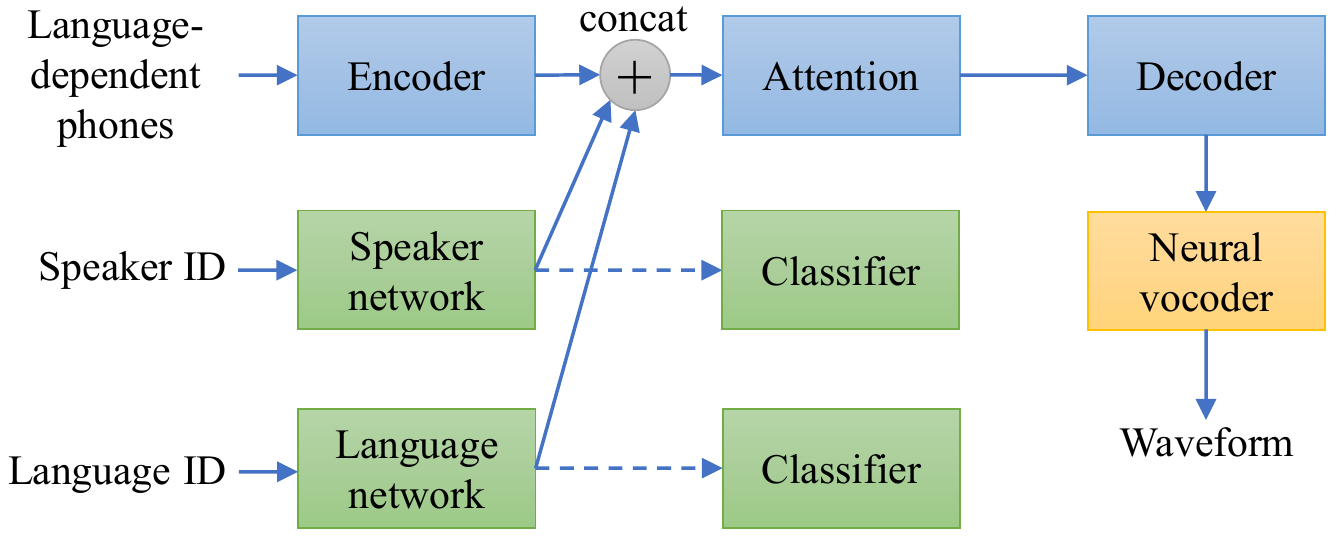}
		\end{overpic}
		\caption{The framework using multi-task learning.}
		\label{fig:MTL}
	\end{center}
\end{figure}

\subsection{Multi-task learning}
\label{sec:MTL}
Multi-task learning can be viewed as a form of inductive transfer. Inductive transfer 
can help improve a model by introducing a learning bias, which causes a model to prefer 
some hypotheses over others~\cite{Ruder:MTL:2017}. In this work, the learning bias is 
introduced by using additional classifiers to encourage the speaker and language 
embeddings to capture the speaker and language characteristics. The MTL framework 
is illustrated in Fig.~\ref{fig:MTL}. 
Based on the multilingual transformer TTS~\cite{Yang:UniTTS:2020}, two classifiers
are introduced, followed by the speaker and language networks. The classifiers are two feed-forward neural networks with softmax outputs, corresponding to the speaker and language identities respectively. In training of the multilingual model using MTL, two additional losses are introduced by the classifiers, i.e. two cross-entropy (CE) losses.


\subsection{Joint training with the speaker classifier}
\label{sec:Joint}
The most direct way to improve the cross-lingual speaker similarity is to
introduce a constrain on the generated cross-lingual speech. In this work,
the x-vector speaker classifier system~\cite{Snyder:Xvec:2018} is introduced. By minimizing 
the distance between the x-vectors of the recording and the cross-lingual
speech for the same speaker, the multilingual model can be updated towards
generating cross-lingual speech with high speaker similarity.
The x-vector system introduces an additional learning bias to the multilingual model, thus joint training with the speaker classifier can also be viewed as a MTL approach.

\begin{figure}[t!]
	\begin{center}
		\begin{overpic}[grid=False,width=6.2cm]{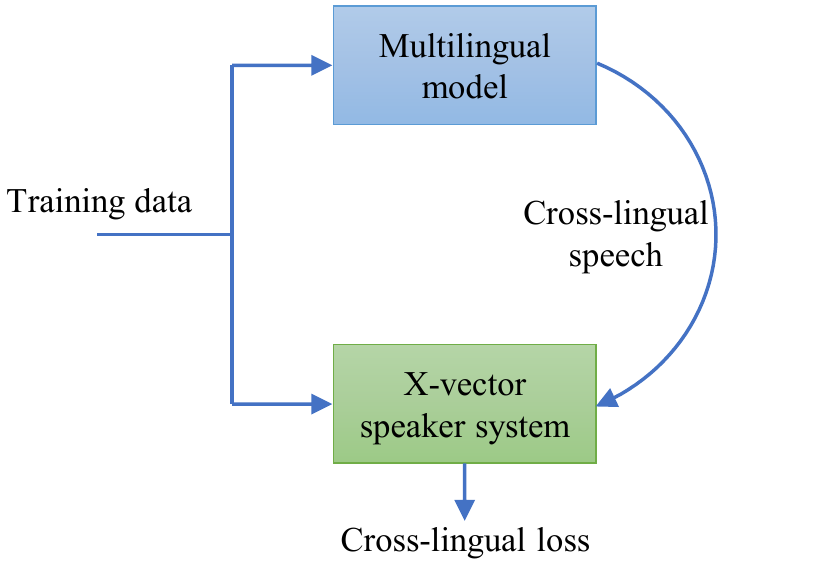}
		\end{overpic}
		\caption{The framework using speaker classifier joint training.}
		\label{fig:Joint}
	\end{center}
\end{figure}

The joint training framework can be summarized in Fig.~\ref{fig:Joint}.
In this figure, the multilingual model can be the vanilla multilingual
transformer TTS or the MTL framework illustrated in Fig.~\ref{fig:MTL}. 
The speaker system can be any network which predicts the speaker 
identity given the input speech. The x-vector system is used in this
work. Compared with the vanilla multilingual model, in joint training with the speaker classifier, two additional losses are
introduced. One is the CE loss for the speaker classifier. Another one is the
cross-lingual loss $\mathcal{L}_{\text{cross}}$ for the multilingual model: 
\begin{flalign} \label{equ:loss}
	\mathcal{L}_{\text{cross}} = \sum_{l} \sum_{l' \neq l} 
	\sum_{\obs_s^{l},\obs_s^{l'}} f_{\text{dist}} 
	\Big( f_{\text{xvec}}(\obs_s^{l}), f_{\text{xvec}}(\obs_s^{l'}) \Big)
\end{flalign}
where $\obs_s^l$ is the training audio sample for speaker $s$ in language $l$, 
and $\obs_s^{l'}$ is the synthesized cross-lingual sample for speaker $s$ in 
language $l'$. In training, the input phone sequence used to generate the cross-lingual
sample $\obs_s^{l'}$ is randomly chosen from the training set of language $l'$.
$f_{\text{xvec}}(\cdot)$ is the x-vector system, which maps variable-length audio samples to fixed-dimensional vectors.
$f_{\text{dist}}(\cdot)$ is a distance function, e.g. the cosine
distance or L2 norm. In this work, L2 norm is used. It is worth noting that, when minimizing the cross-lingual
loss $\mathcal{L}_{\text{cross}}$, only the parameters of the multilingual 
model are updated.
Although this cross-lingual loss targets to minimize the speaker distance between the recordings and the cross-lingual samples, the multilingual model can still keep good speaker similarity in intra-lingual synthesis, as the original mean squared error (MSE) loss between the ground-truth and the predicted Mels is a strong constraint, which can lead to good speaker similarity for the predicted intra-lingual speech.

To minimize the cross-lingual loss $\mathcal{L}_{\text{cross}}$ described in equation \eqref{equ:loss}, the multilingual model 
needs to be operated in inference mode to generate the cross-lingual speech. 
Therefore, teacher forcing cannot be used in training, and the transformer model cannot
be trained in parallel. Moreover, it is impractical to apply back-propagation to an auto-regressive loop for an audio sequence, which usually is long. To alleviate this problem, 
the cross-lingual audio samples can be generated on the fly, then
the generated samples can be treated as the ground truth in the teacher
forcing mode. This approximation is similar to the operation 
used in parallel scheduled sampling~\cite{Duckworth:PSS:2019,Zhou:PSS:2019}.

\section{Experiments}
\label{sec:exp}
\begin{figure}
	\begin{center}
		\begin{overpic}[grid=false,width=8.2cm]{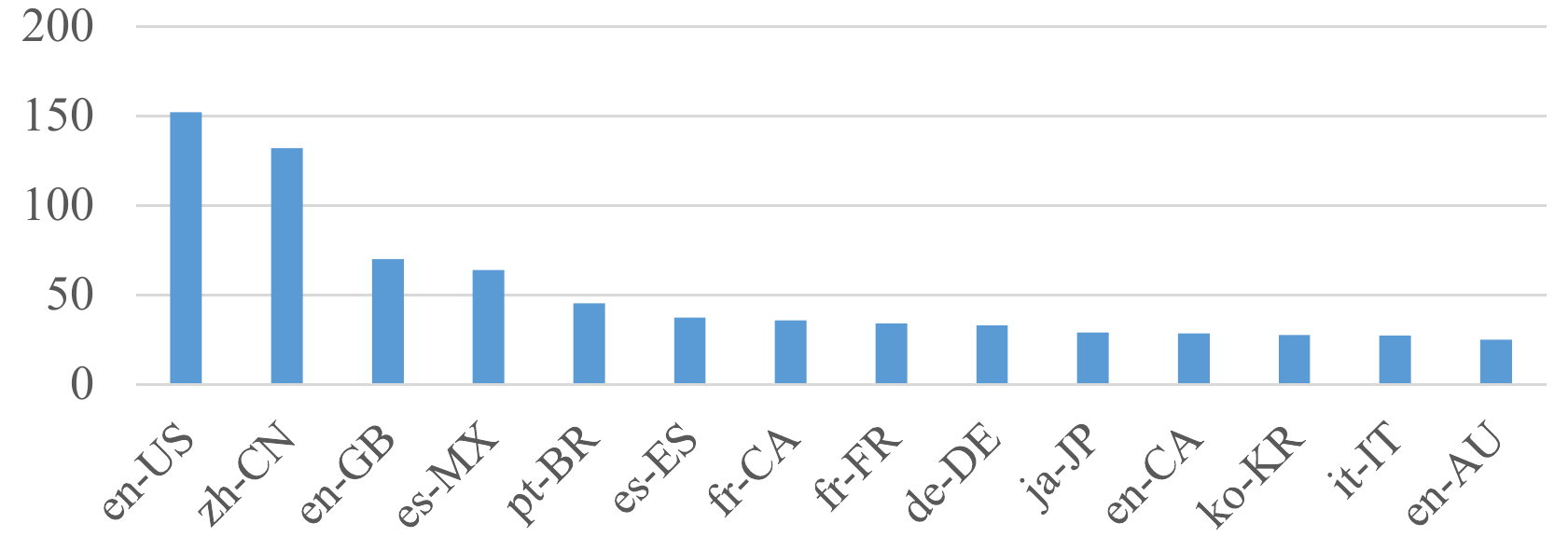}
		\end{overpic}
		\caption{The data distribution over 14 language locales.}
		\label{fig:data_distr}
	\end{center}
\end{figure}

The training
corpus is comprised of around 700 hours professional recordings 
from 14 language locales. In each locale, there are at least 3 speakers.
In this work, the same language from different 
locales is treated independently, i.e. different phone sets and language identities are used.
The amount of training data is unbalanced for different locales, and
the data distribution
over 14 language locales\footnote{The abbreviation for each locale
consists of the language code and the locale ID, e.g. en is English, US is the United States.} is illustrated in Fig.~\ref{fig:data_distr}.
Thus, the language-balanced training strategy discussed in~\cite{Yang:UniTTS:2020}
is used throughout this work.

In training, all the audio samples are down-sampled to 16 kHz, and the beginning 
and ending silences are trimmed to a fixed length, i.e. 30 ms. 
The Adam optimizer 
is used with initial learning rate $10^{-3}$, and exponential decay 
after 100k steps. The minimum learning rate is set to $10^{-5}$.
The baseline multilingual transformer model has the same model structure 
as~\cite{Yang:UniTTS:2020}, with encoder model dimension\footnote{The model dimension indicates the dimensions of the keys, values and queries.} 512, and decoder model dimension 768. Both the speaker and language embedding dimensions are 128.
The universal WaveNet vocoder described in~\cite{Yang:UniTTS:2020} 
is used in waveform generation. Once the vocoder
is trained, it can be applied to the spectrogram from 
any speaker in any language without additional adaptation or fine-tuning procedure.
In the speaker classifier joint training experiments, 
the x-vector system is used, which has the same structure
as~\cite{Snyder:Xvec:2018}, but with a smaller hidden size 256.

When calculating the cross-lingual 
loss $\mathcal{L}_{\text{cross}}$ in equation~\eqref{equ:loss},
it is very computationally expensive to generate the cross-lingual speech $\obs_s^{l'}$, as transformer inference cannot be paralleled.
The sum over all cross-lingual languages $l'$ will make it even worse. In order to make training more efficient, an approximation is made in the experiments. Rather than summing over $l'$, $l'$ is randomly chosen from the training language set except $l$. 
In experiments, we found the cross-lingual loss is a very strong
constrain on the model and it is still slow to train. To avoid overfitting and 
improve training efficiency, the cross-lingual
loss is introduced when both the multilingual model and speaker
classifier model (x-vector system) are fully converged, then jointly update the
whole model with the cross-lingual loss. 
Moreover, the cross-lingual loss is used once every 20 steps. This recipe 
yields a good speaker similarity and an efficient training procedure without 
degrading the voice quality.

\begin{table}[t]
	\begin{center}
		\caption{Cross-lingual evaluations on a zh-CN speaker.}
		\begin{tabular}{ c c c c c }
			\hline
			  \multicolumn{2}{c}{Systems} & Baseline & +MTL & +MTL+Joint \\
			\hline
			\hline
			 \multirow{3}{*}{en-US} & Cos. & 0.173 & 0.097 & \bf 0.067 \\
			 & Sim. & 3.96$\pm$0.08 & 4.02$\pm$0.08 & \bf 4.05$\pm$0.07 \\
			 & Nat. & 3.60$\pm$0.09 & 3.52$\pm$0.08 & 3.56$\pm$0.09 \\
			\hline
			 \multirow{3}{*}{de-DE} & Cos. & 0.246 & 0.113 & \bf 0.076 \\
			 & Sim. & 3.16$\pm$0.10 & 3.84$\pm$0.07 & \bf 3.91$\pm$0.08 \\
			 & Nat. & 2.91$\pm$0.13 & 3.24$\pm$0.12 & 3.16$\pm$0.11 \\
		    \hline
		\end{tabular}
		\label{tab:HJY2}
	\end{center}
\end{table}

\begin{table}[t]
	\begin{center}
		\caption{Cross-lingual evaluations on an en-US speaker.}
		\begin{tabular}{ c c c c c }
			\hline
			  \multicolumn{2}{c}{Systems} & Baseline & +MTL & +MTL+Joint \\
			\hline
			\hline
			 \multirow{3}{*}{zh-CN} & Cos. & 0.291 & 0.234 & \bf 0.188 \\
			 & Sim. & 3.62$\pm$0.09 & 3.67$\pm$0.09 & \bf 3.72$\pm$0.09 \\
			 & Nat. & 3.32$\pm$0.08 & 3.20$\pm$0.09 & 3.19$\pm$0.09 \\
			\hline
			 \multirow{3}{*}{de-DE} & Cos. & 0.255 & 0.231 & \bf 0.227 \\
			 & Sim. & 3.41$\pm$0.11 & 3.47$\pm$0.11 & \bf 3.64$\pm$0.10 \\
			 & Nat. & 3.77$\pm$0.09 & 3.74$\pm$0.09 & 3.89$\pm$0.08 \\
		    \hline
		\end{tabular}
		\label{tab:Jessa2}
	\end{center}
\end{table}

In evaluation, the synthesized audio samples are evaluated by the crowd-sourced subjective listening tests and the objective tests from a speaker classifier. 
In subjective tests, the mean opinion score (MOS) is used to rate 
the cross-lingual naturalness and similarity to the target 
speakers, with range from 1 to 5.
In each similarity evaluation, 15 crowd-sourcing judges are from the source language, and 15 are from the target.
Whereas, in each naturalness evaluation, 20 crowd-sourcing judges are all from the target language.
In objective tests, an independently trained ResNet x-vector speaker classifier system~\cite{Chen:Spk:2020} 
is used to score the speaker similarity.
Cosine distance is used in evaluations, with range from 0 to 2. 
Smaller value indicates better speaker similarity.
This x-vector system is trained with
VoxCeleb 1 and 2 data sets~\cite{Nagrani:VoxCeleb1:2017,Chung:VoxCeleb2:2018}
from 7363 speakers with 2794 hours of speech in
total. The detail model architecture can be found in~\cite{Chen:Spk:2020}.
The audio samples associated with this work are available on this web page\footnote{Audio samples: \href{https://jingy308.github.io/JointSpk}{https://jingy308.github.io/JointSpk}}.

\begin{table}[t]
	\begin{center}
		\caption{Objective similarity evaluations in different languages.}
		\begin{tabular}{ c c c c c c }
			\hline
			  \multicolumn{2}{c}{Languages} & en-GB & es-ES & ja-JP & it-IT \\
			\hline
			\hline
			 \multirow{3}{*}{Intra} & Baseline & 0.134 & 0.091 & 0.142 & 0.115 \\
			 & +MTL & 0.126 & 0.084 & 0.124 & 0.118 \\
			 & +MTL+Joint & \bf 0.125 & \bf 0.080 & \bf 0.117 & \bf 0.102 \\
			\hline
			 \multirow{3}{*}{Cross} & Baseline & 0.241 & 0.163 & 0.183 & 0.152 \\
			 & +MTL & 0.234 & 0.168 & 0.168 & 0.141 \\
			 & +MTL+Joint & \bf 0.185 & \bf 0.146 & \bf 0.134 & \bf 0.125 \\
		    \hline
		\end{tabular}
		\label{tab:obj}
	\end{center}
\end{table}

\subsection{Multi-task learning and joint training}
In cross-lingual synthesis, without using polyglot data or any constrain in training,
the speaker similarity of the generated speech cannot be guaranteed. By using MTL and speaker 
classifier joint training, a learning bias can be introduced to improve the 
speaker representations of the model. In this section, these modified systems are
compared with the baseline model. 
The baseline multilingual transformer model has the same model architecture as~\cite{Yang:UniTTS:2020}. This system is denoted as ``Baseline'' in the experiments.
The MTL system is based on this multilingual transformer by introducing additional 
classification tasks as discussed in section~\ref{sec:MTL}. This system is indicated
by ``+MTL''. The system using speaker classifier
jointing training is based on the MTL system, and an additional x-vector system is introduced 
to be jointly trained with the MTL multilingual system as described in section~\ref{sec:Joint}. 
This system is represented by ``+MTL+Joint''. 
The cross-lingual experiments on the seen speakers during training are tabulated 
in Table~\ref{tab:HJY2} and Table~\ref{tab:Jessa2}.
In the tables, ``Cos.'' denotes the average cosine distance between the recordings and the synthesized 
cross-lingual speech. ``Sim.'' and ``Nat.'' are the similarity and naturalness MOS respectively.
This type of notation is used throughout this paper. The best subjective and objective similarity values are marked in bold.

In Table~\ref{tab:HJY2}, the cross-lingual experiments of the zh-CN (Chinese) speaker are
evaluated in languages en-US (American English) and de-DE (German). By using MTL, the speaker
similarity can be improved in both languages. This improvement can be reflected 
in both the cosine distance and the similarity MOS. The smaller cosine distance represents better speaker
similarity. By introducing speaker classifier joint training, the speaker similarity 
can be further improved, with smaller cosine distance and greater similarity MOS.
For reference, the naturalness MOS to different systems are also given. 
In some languages, MTL improves the naturalness,
such as de-DE for this speaker. In some languages, MTL worsens the naturalness. 
However, in average, MTL does not degrade the naturalness of the cross-lingual speech.
The naturalness scores
between the ``+MTL'' and ``+MTL+Joint'' systems are very close in different languages. Thus,
in general, joint training with speaker classifier does not degrade the naturalness 
of the speech, but increases speaker similarity.

Similar cross-lingual experiments are carried on an en-US speaker as tabulated in Table~\ref{tab:Jessa2}. Cross-lingual
synthesis is evaluated in languages zh-CN and de-DE. The same conclusions can be drawn. 
MTL improves the speaker similarity in terms of both the cosine distance and 
similarity MOS. By introducing speaker classifier joint
training, the similarity can be further improved. The naturalness MOS are also 
given for reference, and speaker classifier joint training does not degrade 
the naturalness of the cross-lingual speech. In the following experiments, the
naturalness scores are not reported.

In the evaluations of the zh-CN and en-US speakers, both the cosine distance and the speaker similarity MOS
are given. The similarity MOS evaluation is relatively heavy, as the judges 
from both the source and target languages are needed. Whereas the cosine 
distance is an objective score, which can be easily computed. The conclusions
from the cosine distances are consistent with the similarity MOS.
In the same experiment, the smaller cosine distance corresponds to a 
greater similarity MOS. Thus, the cosine distance is a good objective metric of 
the speaker similarity. 

In addition to zh-CN and en-US, more comprehensive objective evaluations are carried on other 4 languages with various amount of 
training data, i.e. en-GB (British English),
es-ES (Spanish), ja-JP (Japanese) and it-IT (Italian). 2 speakers are randomly 
chosen in each language with balanced gender in total. Both the intra-lingual 
and cross-lingual cosine distances are tabulated in 
Table~\ref{tab:obj}. In intra-lingual evaluations, the average cosine distances 
are from the 2 speakers in the source language. In cross-lingual evaluations,
the average distances are from the 6 speakers in other 3 languages. According 
to the results in Table~\ref{tab:obj}, By using MTL or speaker classifier joint
training, the overall speaker similarity can be improved. Although MTL does
not yield performance gains in es-ES cross-lingual and it-IT intra-lingual
synthesis, by further using speaker classifier joint training, the similarity 
can be clearly improved. As tabulated in Table~\ref{tab:obj}, the similarity 
improvement in cross-lingual synthesis is greater than that in intra-lingual, 
as the proposed cross-lingual loss targets to improve the cross-lingual 
speaker similarity. In general, by using MTL and speaker classifier joint 
training, the speaker similarity can be consistently improved in each language. 




\begin{table}[t]
	\begin{center}
	    \caption{Cross-lingual evaluations on a new zh-CN speaker.}
		\begin{tabular}{ c c c c c }
			\hline
			  \multicolumn{2}{c}{Systems} & Baseline & +MTL+Joint \\
			\hline
			\hline
			 \multirow{2}{*}{en-US} & Cos. & 0.191 & \bf 0.126 \\
			 & Sim. & 3.99$\pm$0.09 & \bf 4.21$\pm$0.07 \\
			\hline
		\end{tabular}
		\label{tab:Yaya}
	\end{center}
\end{table}

\begin{table}[t]
	\begin{center}
	    \caption{Objective evaluations on a new zh-CN speaker.}
		\begin{tabular}{ c c c c c c }
			\hline
			  Languages & zh-CN & es-ES & ja-JP & it-IT \\
			\hline
			\hline
			 Baseline & 0.096 & 0.209 & 0.105 & 0.216 \\
			 +MTL+Joint & \bf 0.092 & \bf 0.126 & \bf 0.093 & \bf 0.101 \\
			\hline
		\end{tabular}
		\vspace{-2mm}
		\label{tab:Yaya_obj}
	\end{center}
\end{table}

\subsection{Experiments on new speakers}
The multilingual model is trained with a limited number of speakers.
Therefore, it is common to extend the model to a new speaker, which is 
unseen in the training set, with a limited amount of data.
When extending to new speakers,
the whole multilingual model is refined. To avoid overfitting 
to the target speaker, the data from the new speaker and 
the existing speakers are used in model refining.
Normally, the amount of data from the new speaker is very limited, 
whereas the data amount from other speakers is very huge. Thus, 
language-balanced training described in~\cite{Yang:UniTTS:2020}
is also used in speaker extension. 
The effect of MTL can be clearly observed from previous experiments, 
thus in the following experiments, only the final ``+MTL+Joint'' model
is compared with the baseline model.

Table~\ref{tab:Yaya} tabulates the cross-lingual evaluation results of 
a new zh-CN speaker. There are around 9 minutes of data available for 
training. By using MTL and speaker classifier joint training,
smaller cosine distance and greater similarity MOS can be achieved.
This means an increase in the speaker similarity between the cross-lingual 
speech and the original native recordings of this new speaker.
More objective evaluation results are given in Table~\ref{tab:Yaya_obj}.
The proposed model yields consistent similarity gains in each language,
and in general the gains in cross-lingual synthesis are much greater 
than that in intra-lingual.
Similar experiment is carried on a new en-GB (British English) speaker, 
with around 5 minutes of data. The experimental results are tabulated in
Table~\ref{tab:Gordon} and \ref{tab:Gordon_obj}. The same conclusion 
can be drawn on this speaker in both subjective and objective evaluations.
By using MTL and speaker classifier joint training, the 
speaker similarity to this new speaker can be improved, with greater similarity MOS and smaller cosine distance in each language.

\comment{
In all experiments, both the cosine distance and the speaker similarity MOS
are given. The similarity MOS evaluation is relatively heavy, as the judges 
from both the source and target languages are needed. Whereas the cosine 
distance is an objective score, which can be easily computed. The experiments show
the cosine distance is negatively correlated with the similarity MOS.
In the same experiment, the smaller cosine distance corresponds to a 
greater similarity MOS. Thus, the cosine distance is a good measure of 
the speaker similarity, and can be easily evaluated. 
}

\begin{table}[t]
	\begin{center}
	    \caption{Cross-lingual evaluations on a new en-GB speaker.}
		\begin{tabular}{ c c c c c }
			\hline
			  \multicolumn{2}{c}{Systems} & Baseline & +MTL+Joint \\
			\hline
			\hline
			 \multirow{2}{*}{zh-CN} & Cos. & 0.198 & \bf 0.194 \\
			 & Sim. & 3.57$\pm$0.09 & \bf 3.68$\pm$0.09 \\
			\hline
		\end{tabular}
		\label{tab:Gordon}
	\end{center}
\end{table}

\begin{table}[t]
	\begin{center}
	    \caption{Objective evaluations on a new en-GB speaker.}
		\begin{tabular}{ c c c c c c }
			\hline
			  Languages & en-GB & es-ES & ja-JP & it-IT \\
			\hline
			\hline
			 Baseline & 0.088 & 0.219 & 0.198 & 0.224 \\
			 +MTL+Joint & \bf 0.083 & \bf 0.202 & \bf 0.188 & \bf 0.164 \\
			\hline
		\end{tabular}
		\vspace{-2mm}
		\label{tab:Gordon_obj}
	\end{center}
\end{table}

\section{Conclusions and the future work}
\label{sec:conc}

This paper studies the approaches to improve the speaker similarity
for the multilingual transformer TTS model. By using multi-task learning, 
the overall speaker similarity for the model can be improved.
Joint training with an x-vector system, the speaker similarity can
be further improved. 
These techniques can also be easily extended to model new speakers, and 
the speaker similarity to new speakers can be improved as well.
Moreover, our experiments show that the cosine distances from the x-vector system are consistent with the similarity MOS, and this objective score
can be easily computed. 
The naturalness of the cross-lingual speech is still a drawback for 
the multilingual model. The future work will focus on the improvement 
in the voice quality and naturalness of the cross-lingual speech.

\section{Acknowledgments}
\label{sec:acknoledge}
The authors would like to thank Yan Deng, Wenning Wei, Xi Wang and Liping Chen for their help in this work.

\bibliographystyle{IEEEtran}

\bibliography{mybib}


\end{document}

%% file: Notations.tex
\newcommand{\obs}{\boldsymbol{o}}

